\documentclass[twocolumn]{aastex63}

\usepackage{graphicx}

\shorttitle{Protoplanetary Disks in Serpens}
\shortauthors{Anderson et al.}

\graphicspath{{./}{figures/}}

\begin{document}

\title{Protostellar and Protoplanetary Disk Masses in the Serpens Region}

\author[0000-0002-4876-630X]{Alexa R. Anderson}
\affiliation{Institute for Astronomy, University of Hawai'i at Mānoa, 2680 Woodlawn Drive, Honolulu, HI 96822, USA}

\author[0000-0001-5058-695X]{Jonathan P. Williams}
\affiliation{Institute for Astronomy, University of Hawai'i at Mānoa, 2680 Woodlawn Drive, Honolulu, HI 96822, USA}

\author[0000-0003-2458-9756]{Nienke van der Marel}
\affiliation{Department of Physics \& Astronomy, University of Victoria, 3800 Finnerty Road, Victoria, BC V8P 5C2, Canada}

\author[0000-0003-1413-1776]{Charles J. Law}
\affiliation{Center for Astrophysics \textbar\, Harvard \& Smithsonian, 60 Garden St., Cambridge, MA 02138, USA}

\author{Luca Ricci}
\affiliation{Department of Physics and Astronomy, California State University Northridge, 18111 Nordhoff Street, Northridge, CA 91330, USA}

\author[0000-0002-6195-0152]{John J. Tobin}
\affiliation{National Radio Astronomy Observatory, 520 Edgemont Rd., Charlottesville, VA 22903, USA}

\author[0000-0001-5229-2402]{Simin Tong}
\affiliation{Leiden Observatory, Niels Bohrweg 2, 2333 CA Leiden, The Netherlands}

\begin{abstract}
    We present the results from an Atacama Large Millimeter/Submillimeter Array (ALMA) 1.3 mm continuum and $^{12}$CO ($J=2-1$) line survey spread over 10 square degrees in the Serpens star-forming region of 320 young stellar objects, 302 of which are likely members of Serpens (16 Class I, 35 Flat spectrum, 235 Class II, and 16 Class III). From the continuum data, we derive disk dust masses and show that they systematically decline from Class I to Flat spectrum to Class II sources. Grouped by stellar evolutionary state, the disk mass distributions are similar to other young ($<3$\,Myr) regions, indicating that the large scale environment of a star-forming region does not strongly affect its overall disk dust mass properties. These comparisons between populations reinforce previous conclusions that disks in the Ophiuchus star-forming region have anomalously low masses at all evolutionary stages. Additionally, we find a single deeply embedded protostar that has not been documented elsewhere in the literature and, from the CO line data, 15 protostellar outflows which we catalog here.
\end{abstract}


\section{Introduction}

Stars are born from cores within dusty molecular clouds. Though it is widely accepted that planets form in circumstellar disks surrounding these young stars, the effect of disk properties on planet formation is still relatively unknown. Especially uncertain, but important to identify, are the effects of the large cloud scale environment such as stellar density and strong radiation fields, in regulating star, disk, and planet formation.

Most stars, and therefore their accompanying protoplanetary disks, form in clusters \citep{Lada_2003}. If larger than $\sim 200$ members, these clusters generally contain at least one OB star \citep{McKee_1997} and their accompanying strong UV radiation field enhances photoevaporation and may decrease the dust masses of disks nearby \citep{Mann_2014, Ansdell_2017}, depleting the reservoir of planet-forming material. However, dynamical interactions in a dense cluster can also reduce disk masses and lifetimes \citep{Bate_2018}. To distinguish between photoevaporative \citep{Rosotti_2014} and dynamical effects requires the study of a young, dense star-forming region that lacks massive stars. The Serpens star-forming region \citep{Gutermuth_2008b} is one such natural laboratory.

The high visual extinction and low Galactic latitude toward the Serpens and Aquila regions (RA$\sim$18h, Dec$\sim$00$^{\circ}$) led to prior confusion over membership and distance. With the advent of \textit{Gaia} \citep{Gaia_2016, Gaia_2021}, it has become clear that most young stellar objects (YSOs) in Serpens are clustered into four main groups in a common star-forming region spread over roughly 10 square degrees at a distance ranging from $\sim 400-500$\,pc \citep{Herczeg_2019, Dzib_2010, Ortiz-Leon_2017, Ortiz_Le_n_2018, Zucker_2019}. The tight clustering, associated cloud, and large numbers of protostars all indicate a young, relatively unevolved region.

Multiple studies have observed nearby young star-forming regions in an effort to characterize the statistical properties of protoplanetary disk populations. In doing so, these surveys provide constraints on the initial conditions of planet formation. Like Serpens, young star-forming regions  ($<3$\,Myr) have been the primary target of such studies (Lupus, Taurus,  Orion, Perseus, Ophiuchus; \citealp{Andrews_2013, Ansdell_2016, Tychoniec_2020, Tobin_2020, Cieza_2019}). These works have predominantly found that (sub-)millimeter luminosities decrease with time due to dust processing whereas the gas and micron-dust may dissipate separately on longer timescale \citep{Michel_2021}. However, it is necessary to identify if there are secondary dependencies, in particular the presence of massive stars and high stellar density.

A previous survey of the Serpens region was completed by \citet{Law_2017}. They used the Submillimeter Array (SMA) to observe 62 disks at a resolution of $\sim3''$. 
Here we use ALMA to observe the same region to increase the sample size of disks in the region by a factor of more than 5 with greater sensitivity. The primary goal of this project is to analyze the distribution of disk masses within Serpens and compare it to other nearby star forming regions. The layout of the paper is as follows: we describe the survey sample in \S\ref{sec:sample}, the ALMA observations in \S\ref{sec:obs}, and our results in \S\ref{sec:results}. The mass distributions with comparisons to other regions are depicted in \S\ref{sec:distributions}. We discuss the implications of the work in \S\ref{sec:discussion} and summarizing our findings in \S\ref{sec:summary}, while the protostellar outflows seen in the CO line are discussed in Appendix \ref{sec:outflows}.

\section{Source selection and classification}
\label{sec:sample}

\begin{figure*}[htb!]
    \centering
    \includegraphics{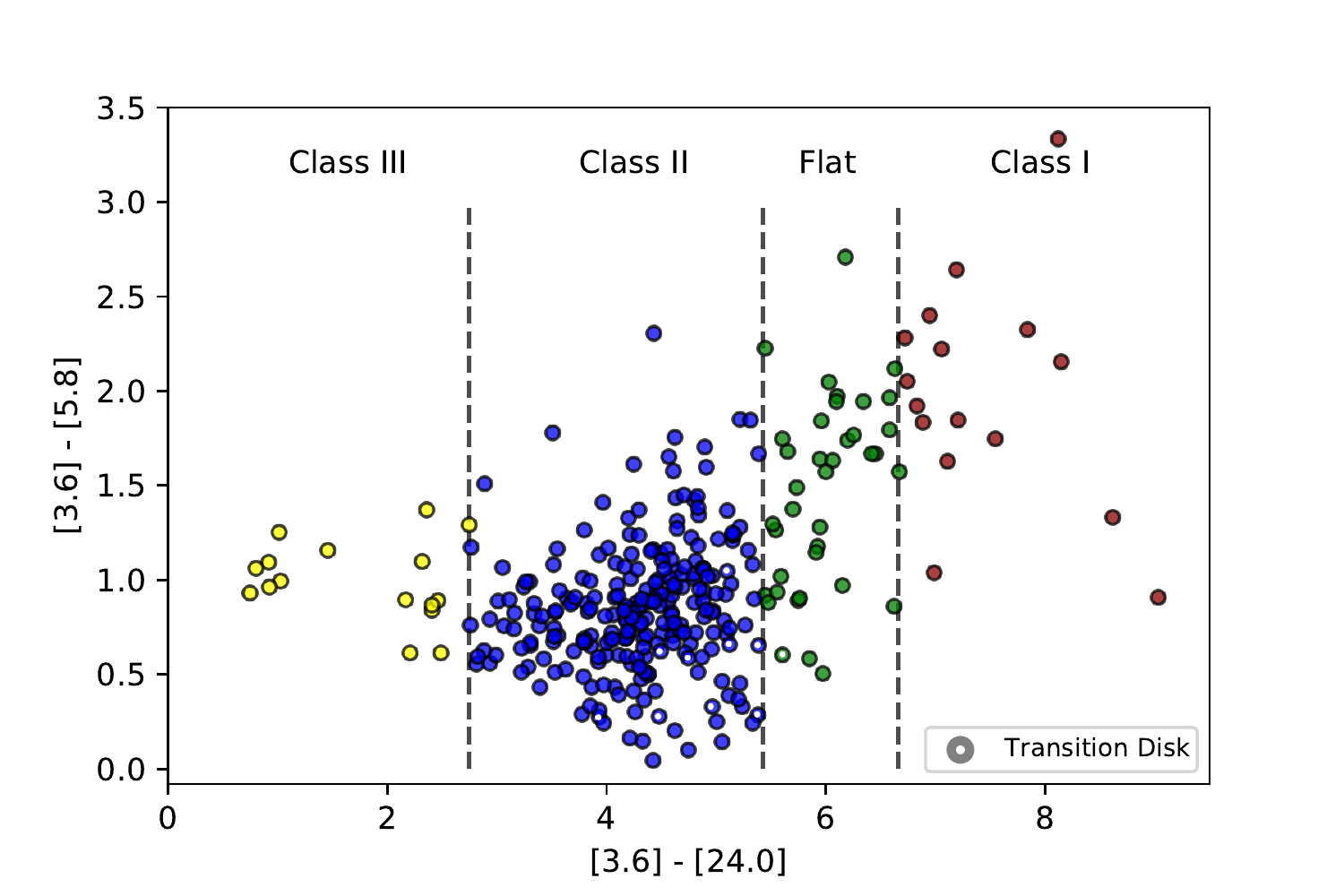}
    \caption{Spitzer IRAC [3.6] - [24.0] versus [3.6] - [5.8] color-color diagram used for classifying YSOs. The dashed black lines in this magnitude plot correspond to the $\alpha_{IR}$ slopes that we use to delineate objects into their evolutionary classes (Spitzer IRAC [3.6] - [24.0] colors of 2.75, 5.42, and 6.66). Transition disk candidates are indicated by the points with white circles inset.
    }
    \label{fig: color-color}
\end{figure*}

\begin{figure}[htb!]
    \centering
    \includegraphics[width=\columnwidth]{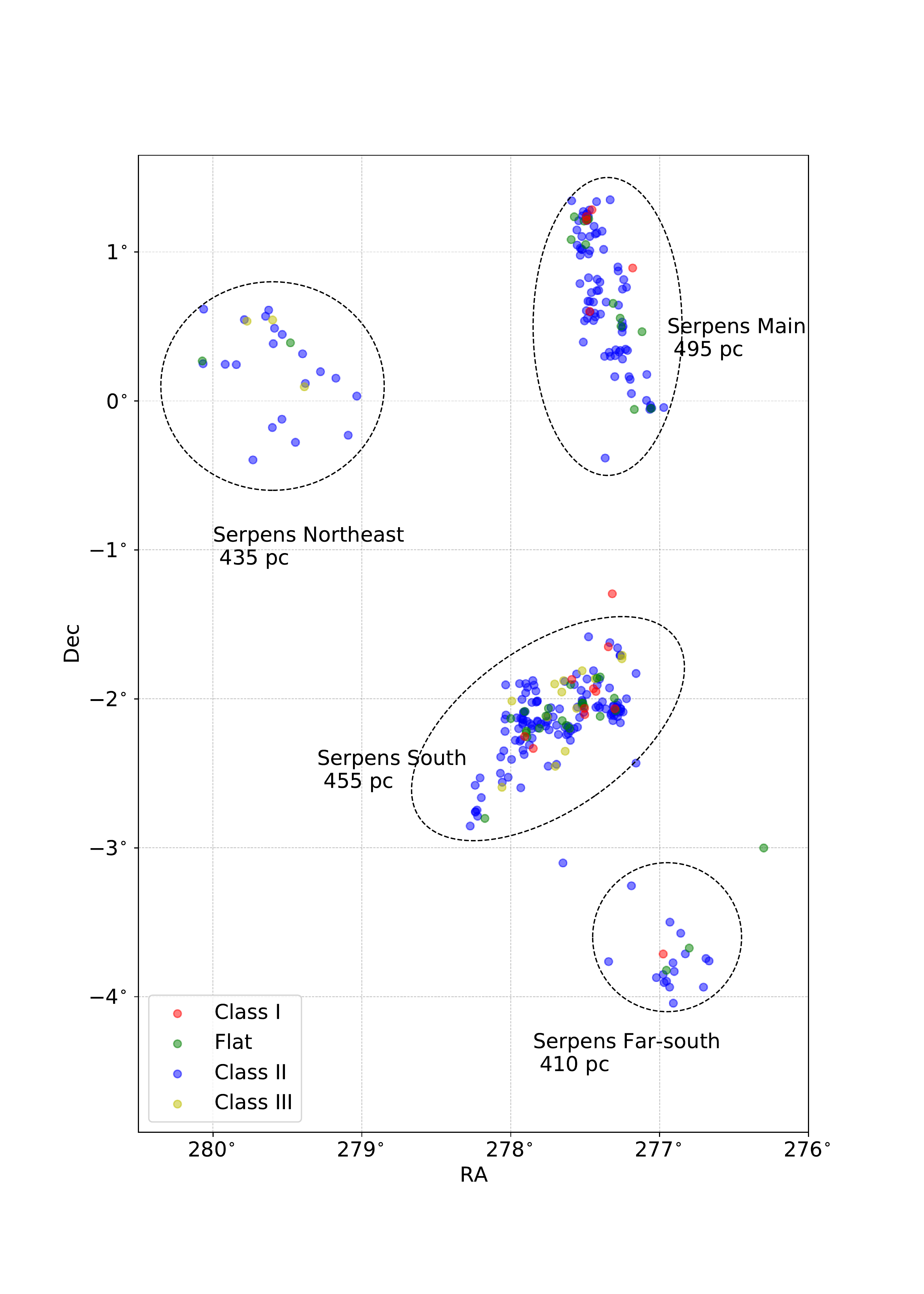}
    \caption{Location of observed disks in the Serpens region.
    Four different groups are visually identified and labeled as Serpens Northeast, Serpens Main, Serpens South, and Serpens far-south, each with the mean distances from \textit{Gaia} EDR3 parallaxes. An additional two sources are present in our sample, but reside at an RA of nearly 270$^\circ$, and so are not shown here.}
    \label{fig: distances}
\end{figure}

The Spitzer c2d and Gould Belt surveys identified 1546 candidate YSOs toward the Serpens region \citep{Dunham_2015}. We reduced this to a sample size of 320 for our ALMA survey by selecting only those sources with infrared spectral slopes $<$ 1.6, as defined by 
\begin{equation}
    \alpha_{\rm IR} = \frac{d\log\nu F_\nu}{d\log\nu}
\end{equation}
from 2 to 24\,$\mu$m, corresponding to Class I, Flat Spectrum, and Class II YSOs \citep{Williams_2011}; a K-band cutoff of 13\,mag to remove very low mass stars, $\lesssim0.2\,M_\odot$; and a \textit{Gaia DR2}-based distance cutoff of 650\,pc.

The visual extinction toward Serpens is high. Over 50\% of stars toward this region have $A_{\rm V}>7$\,mag \citep{Harvey_2007b} and  Spitzer surveys were restricted to regions of high extinction ($A_{\rm V} > 6$\,mag) to limit confusion from the Galactic background \citep{Dunham_2015}. Consequently, the K-band magnitude is significantly extincted which decreases the spectral slope used for YSO classification.
In general, the difference between the observed and intrinsic slopes between two wavelengths $\lambda_1$ and $\lambda_2$ is
\begin{equation}
    \Delta\alpha_{\lambda_1-\lambda_2}
= \alpha^{\text{obs}} - \alpha^{\text{intrinsic}}
= \frac{A_{\lambda_1}}{2.5\log_{10}(\lambda_1/\lambda_2)}
\end{equation}
where $A_{\lambda}$ is the extinction at a given wavelength and we assume that $\lambda_2$ is large enough that $A_{\lambda_2}=0$.
From the extinction law in \citet{Wang_2019}, when measuring the slope between the 2MASS K band and MIPS $24\,\mu$m, $\Delta\alpha_{\rm IR} \geq$ 0.19 for our sample which is sufficient to shift many sources from Class III to Class II, Class II to Flat, and so on.
We therefore revised the YSO classification to use the slope between IRAC $3.6\,\mu$m and MIPS $24\,\mu$m bands rather than the 2MASS K and MIPS $24\,\mu$m bands. In this case, $\Delta\alpha_{\rm IR}=0.11$ for $A_{\rm V}=6$\,mag, which is small enough to substantially reduce the number of potentially misclassified YSOs.

Though \citet{Dunham_2015} accounted for extinction in constructing SEDs to calculate $\alpha_{\rm{}IR}$ in their original identification of YSO candidates, we choose to base our classification system on the $3.6-24\,\mu$m slope for two main reasons. First, the extinction correction is dependent on an object's intrinsic color, as well as the nature of material along the line of sight and many objects in Serpens have inaccurately determined spectral types, particularly the more embedded ones. Additionally, they assumed a uniform extinction law, though dust properties may vary throughout the region, again especially for the more embedded objects. Our method, using a sightly longer anchor wavelength is less sensitive to YSO temperature and dust properties. Nevertheless, the effect is not large and our classification agrees with \citet{Dunham_2015} for 85\% of the sample.

Figure \ref{fig: color-color} shows the distribution of our objects in color-space grouped by class to illustrate how evolutionary classification is derived from colors. We then cross-correlated our sample with the list of transition disk candidates (TDs) identified by mid-infrared dips in their spectral energy distributions by \citet{vanderMarel_2016}. These disks may have large central gaps or cavities and are demarcated with circles in Figure \ref{fig: color-color}. It is possible that our sample is not complete, particularly for embedded YSOs. However, our sample has similar demographics to other regions presented in \citet{Dunham_2015} and can therefore be considered representative.

In the time between proposal, observations, and analysis, the \textit{Gaia} Early Data Release 3 (\textit{EDR3}) catalog was released. This decreased distance uncertainties and increased the number of sources with parallax measurements. Four sub-regions were identified by visually identifying groups of sources based on their angular coordinates and then confirmed by checking that most sources in these clusters had similar parallaxes. Figure \ref{fig: distances} shows the location of each observed source on the sky. \citet{Herczeg_2019} refer to these clusters as Serpens Northeast, Serpens Main, Serpens South, and Serpens far-south. The four regions are labeled in Figure \ref{fig: distances}, with distances of roughly 435 $\pm$ 25 pc, 495 $\pm$ 90 pc, 455 $\pm$ 50 pc, and 410 $\pm$ 35 pc, respectively.
 We calculated the weighted average of the parallaxes of objects that had \textit{Gaia EDR3} measurements within a subregion and converted this to an average distance. This averaged distance was assigned to objects without parallax measurements. The stellar density is high in these subregions. The densest is Serpens South where the median nearest neighbor distance is 0.1\,pc \citep[][accounting for the revised distance]{Gutermuth_2008b} which is comparable to protostellar core size scales.

Serpens also lies near the HII region W40, and the two are estimated to reside at similar distances \citep{Ortiz_Le_n_2018}. Previous works have confirmed that the two star-forming regions are physically connected and interacting \citep{Shimoikura_2020}. Particularly for objects in Serpens South, star and disk formation may be affected by close proximity to this other stellar nursery.

\section{Observations}
\label{sec:obs}
We observed the sample of 320 targets in ALMA Cycle 7 program 2019.1.00218.S (PI: van der Marel). The observations were carried out in Band 6 in the C-2 configuration (baselines from 15 to 314 m) with 43 antennas on 2019 December $19^{\rm th}$ and $31^{\rm st}$. Each source was observed in a single pointing with an integration time of 20\,s. The precipitable water vapor ranged from 1.0 to 4.1 mm over the course of the observations and the average system temperature was 100 K. Three wide spectral windows provided a total continuum bandwidth of 5.6 GHz centered on 241.4\,GHz and the fourth was a narrow window centered  around the $^{12}$CO 2--1 line (230.536 GHz) at 0.32\,km\,s$^{-1}$\ spectral resolution.

\begin{figure*}[htb!]
    \centering
    \includegraphics[width=0.9\textwidth]{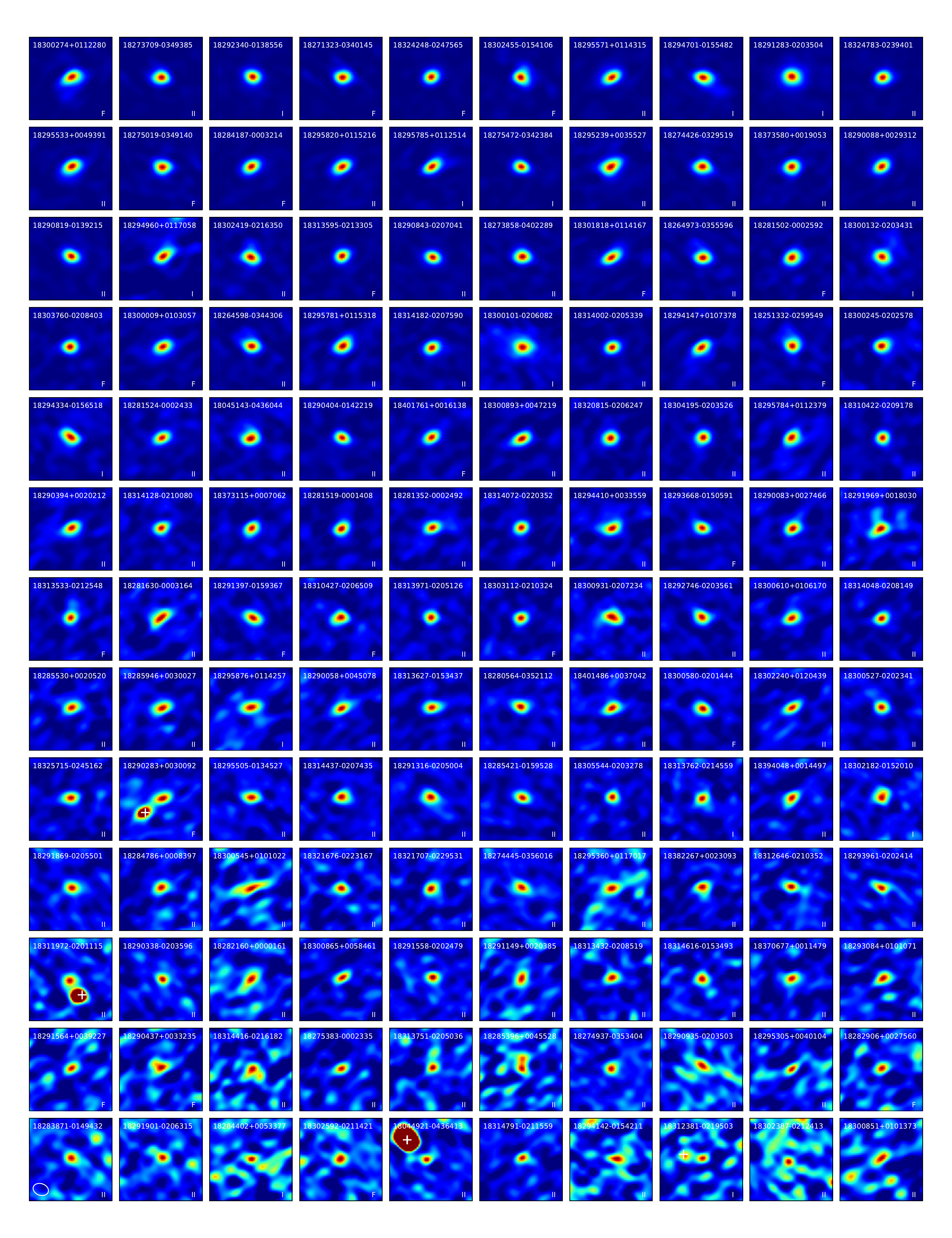}
    \caption{1.3 mm continuum images of the the 130 disks detected in our ALMA Cycle 7 program, ordered by decreasing continuum flux density (as reported in Table \ref{table: props}). Each panel is $7''$ by $7''$, and is scaled from -0.3 mJy to the maximum pixel value of the primary. This causes a few bright secondary sources to be saturated in some fields. White crosses are plotted at the location of maximum intensity for binary companions (defined in \S\ref{sec:results}). A representative beam  size is shown in the lower left panel, and the infrared spectral classification for all sources is noted in the bottom right corner of each panel.}
    \label{fig: serpens all 1}
\end{figure*}

We applied standard imaging techniques to the pipeline calibrated visibilities using the Common Astronomy Software Applications package (CASA, \citealp{McMullin_2007}) {\tt tclean} task. The Briggs robust weighting parameter was set to 0.5, and the beam size ranged from $1\farcs 2 \times 1\farcs0$ to $1\farcs4 \times 1\farcs0$ with a position angle from $-60^{\circ}$ to $75^{\circ}$. The average root-mean-square (rms) uncertainty of the continuum images was 0.45 mJy beam$^{-1}$ but ranged from 0.13 to 3.2 mJy beam$^{-1}$ where the highest values were due to clustering (multiple 3$\sigma$ detections in a single pointing, increasing noise in the image) and low image quality associated with the limited uv coverage in the short integrations. We detected 130 sources in the continuum and show a montage of their images in Figure \ref{fig: serpens all 1}. All are point sources at the observed resolution and the $7''\times 7''$ panels show binary companions in several cases.

The same standard imaging techniques were used to produce channel maps of $^{12}$CO once the continuum had been subtracted from the data with {\tt uvcontsub} from CASA. Each object's channel maps were visually inspected to determine if they contained any emission. These channels were used in the creation of integrated intensity maps for each object. If, for an object, all channels were emission-free, all channels (steps of 0.5 km s$^{-1}$) between -10 and 30 km s$^{-1}$ (relative to the CO rest frame at 230.536 GHz) were used. Then, the CASA task {\tt immoments} was used to create moment 0 integrated intensity maps. A mask was applied to these images such that only pixels with values greater than 50 mJy beam$^{-1}$channel$^{-1}$ were included in the maps. The average rms was $\sim$30 mJy beam$^{-1}$channel$^{-1}$. In general, our sensitivity is insufficient to detect emission from the disk but we were able to identify outflow emission in several cases that we discuss in Section \ref{sec:outflows}. 

\section{Results}
\label{sec:results}

Continuum flux densities were measured using aperture photometry, as in \citet{Ansdell_2016}. Circular apertures were centered on the brightest pixel associated with our target. For brighter sources with flux densities $>$ 5 mJy, we used an aperture radius of $2''$ to ensure most of the emission was captured, and for the fainter sources, a radius of 1$\farcs$5. We chose these on the basis of a curve-of-growth analysis that showed this was where the highest signal-to-noise ratio was typically achieved and also that less than $\sim$10\% of an object's flux extended past either of these aperture boundaries. The rms in the measurement was estimated to be the standard deviation of measurements from ten apertures of the same size uniformly distributed in position angle and spaced two times the aperture radius from the central source. In the case of millimeter binaries, apertures that contained any companion sources were removed from flux calculations, though fluxes of the companions are reported later on in Table \ref{tab: binaries}.

Disks were marked as detections if they were within 1$\arcsec$ of the source's 2MASS coordinates and if the measured flux density was greater than three times the rms. For a handful of cases which were either highly clustered or where the image was affected by a bright off-center source, we measured fluxes manually using the CASA {\tt imview} task with an aperture of 1$\arcsec$. For single sources, we verified our measurements with those measured as a point source in the visibility plane through {\tt uvmodelfit}. Two marginal cases were identified as visual detections, and though they were not counted as such with aperture photometry due to the high degree of clustering in the field, they were confirmed with {\tt uvmodelfit}. Therefore, these cases use the {\tt uvmodelfit} results in our tables and subsequent analysis.

The dust masses of the disks were calculated using the simplest assumptions: optically thin emission and constant dust temperature, so as to allow direct comparisons with other regions (essentially, millimeter luminosity). We use a dust opacity coefficient, $\kappa_{\nu} = (\nu/100\text{GHz})$ cm$^2$ g$^{-1}$
\citep{Beckwith_1990} and, for Class II sources, a dust temperature $T_{\rm dust} = 20$ K.
Then, using the standard formula for a disk with flux density $F_{\nu}$, the dust mass is
\begin{equation}
\label{eq:Mdust}
M_{\rm dust} = \frac{F_{\nu} d^2}{\kappa_{\nu} B_{\nu}(T_{\rm dust})}
= 4.00\,M_\oplus \left(\frac{F_{1.3 \text{mm}}}{1 \text{mJy}}\right) \left(\frac{d}{400 \text{pc}}\right)^2,
\end{equation}
where $B_{\nu}$ is the Planck function,
$\kappa_\nu = 2.41$ cm$^2$ g$^{-1}$ at our observing frequency,
and $d$ is the distance to each source. This approach was the same as that used for Lupus and Ophiuchus Class II disks \citep[e.g.,][]{Ansdell_2016, Williams_2019}.

Protostellar envelopes keep the disk warmer than in Class II sources \citep{vantHoff_2020}. In a survey of embedded Orion objects, \citet{Tobin_2020} derived disk masses using $T_{\rm dust}=40$\,K for solar luminosity Class I sources. For Flat spectrum sources with a thinner envelope, \citet{Tychoniec_2020} used $T_{\rm dust}=30$\,K. We will compare with these works later and use these characteristic dust temperatures for the flux-to-mass conversion. This gives scaling factors of 1.71 (Class I) and 2.40 (Flat) instead of 4.00 in equation~\ref{eq:Mdust}. 
In all cases, non-detections were assigned a mass upper limit corresponding to three times the flux density rms. The median mass limit for the non-detections is $7.6\,M_\oplus$ with a fairly large range (factor of 3) due to the variation in intrinsic noise and, especially, distance across our sample. Unsurprisingly given the sensitivity of our survey, short integration times, and the fact that Class III sources are generally very faint \citep{Lovell_2020, Michel_2021}, all 16 Class III sources were undetected.

\begin{deluxetable*}{ccccccccccccc}[htb!]
\tablecaption{Source parameters\tablenotemark{a}}
\label{table: props}
\tablehead{\colhead{2MASS ID} & \colhead{RA} & \colhead{Dec} & \colhead{$d$} & \colhead{$\sigma_d$} & \colhead{Class} & \colhead{$F_{1.3mm}$} & \colhead{$\sigma_{1.3mm}$} & \colhead{Disk Mass} & \colhead{$\sigma_{\rm{}Mass}$} & \colhead{$F_{^{12}CO}$} & \colhead{$\sigma_{^{12}CO}$} & \colhead{Flag\tablenotemark{b}}\\
\colhead{} & \colhead{$^{\circ}$} & \colhead{$^{\circ}$} & \colhead{pc} & \colhead{pc} & \colhead{} & \colhead{mJy} & \colhead{mJy} & \colhead{$M_{\Earth}$} & \colhead{$M_{\Earth}$} & \colhead{Jy km s$^{-1}$} & \colhead{Jy km s$^{-1}$}}

\startdata
18300274+0112280 & 277.51138 & 1.20784 & 435.73 & 24.05 & F & 68.28 & 3.04 & 194.54 & 23.16 & 5.49 & 0.98 &  \\
18273709-0349385 & 276.90457 & -3.8274 & 379.74 & 16.45 & II & 61.48 & 0.67 & 221.53 & 19.35 & 0.42 & 0.08 &  \\
18292340-0138556 & 277.34753 & -1.64879 & 456.17 & 48.79 & I & 56.38 & 0.44 & 125.46 & 26.86 & 15.81 & 2.52 & O \\
18271323-0340145 & 276.80514 & -3.67071 & 259.51 & 29.92 & F & 44.28 & 2.53 & 82.25 & 17.7 & 0.73 & 0.12 & G \\
18324248-0247565 & 278.17700 & -2.79903 & 456.17 & 48.79 & F & 41.15 & 0.81 & 128.5 & 27.61 & 2.94 & 0.68 & O \\
\enddata
\tablenotetext{a}{The full table is available online in machine readable form.}
\tablenotetext{b}{O: Target is associated with $^{12}$CO outflow\\
B: Target is within 2\farcs5 ($\sim$1000 AU) of another sub-mm detection of at least 3$\sigma$\\
G: Target is located at a very large ($>675$ pc) or small ($<275$ pc) distance\\
T: Object is a transition disk candidate as identified by \citet{vanderMarel_2016}}
\end{deluxetable*}

Table \ref{table: props} lists the basic properties of our target list, including the derived dust masses or upper limits. Flags indicate if the sources have potentially interesting features. O is used for objects that have an associated $^{12}$CO molecular outflow. B is used for objects in binary systems -- a 3$\sigma$ mm detection within 2\farcs5 ($\sim$1000 AU) of the central source. G is used to indicate sources located at distances outside of the probable range of Serpens members. In the change from \textit{Gaia DR2} to \textit{EDR3}, 13 sources were found to lie at very large ($>675$ pc) or small ($<275$ pc) distances -- well outside the boundaries of the cloud. We exclude these in the analysis of the Serpens mass distributions here but include them in the table of results for completeness. T is used for transition disk candidates as identified by \citet{vanderMarel_2016}. Once these sources had been excluded, our sample comprised 16 Class I, 35 Flat spectrum, 235 Class II, and 16 Class III sources.

\subsection{Multiplicity}

Of the 320 observed sources, four were noted to have binary companions (a 3$\sigma$ sub-mm detection within 2\farcs5 of the target source). Because of their small angular separation on the sky and strong sub-mm emission, we assume that they are binary companions at the same distance as our target sources. These objects don't have counterparts in {\it Gaia}, but are visible in optical or infrared regimes in published data. Table \ref{tab: binaries} shows the basic properties of these objects. These objects had not previously been reported as binaries in the literature, and are noted here.

\begin{table}[htb!]
\centering
\caption{Source parameters for binary companions}
\begin{tabular}{ccccc}
\hline \hline
Central Source & $\Delta$RA & $\Delta$Dec & $F_{1.3}$ & $\sigma_{F_{1.3}}$\\
 & $''$ & $''$ & mJy & mJy \\
\hline
18044921-0436413 & 2.4 & 2.0 & 15.11 & 0.51 \\
18290283+0030092 & 1.6 & -1.4 & 5.53 & 0.84 \\
18311972-0201115 & -1.2 & -1.4 & 3.72 & 0.25 \\
18312381-0219503 & 0.8 & 1.0 & 0.88 & 0.28 \\
\hline
\end{tabular}
\label{tab: binaries}
\end{table}

\begin{figure*}
    \includegraphics[width=\textwidth]{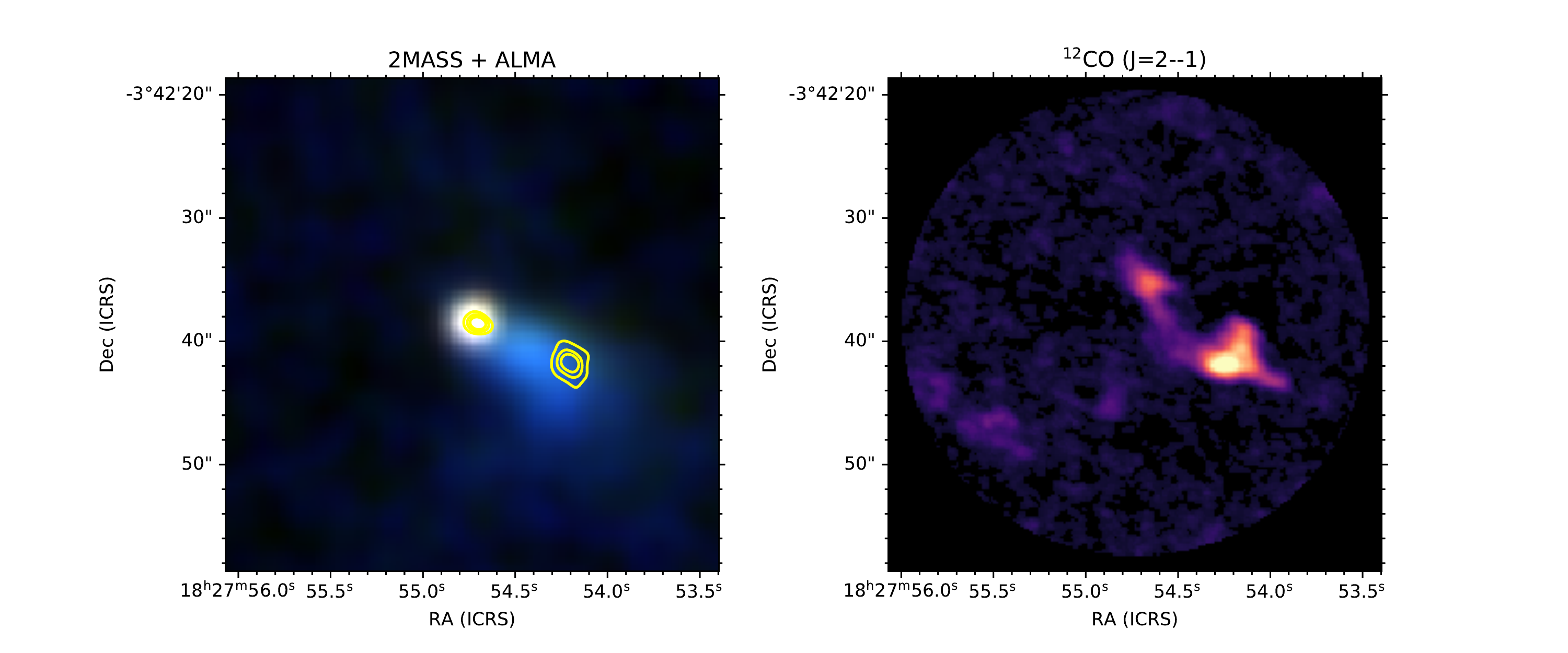}
    \caption{J18275472-0342384 and a nearby Class 0 source. Left: The background image is a three-color composite of the 2MASS J, H, and K filters. Yellow contours represent the 1.3 mm continuum. The central source is illuminated in both NIR and continuum, and a binary companion is identified in the continuum in a portion of the NIR image that contains significant nebulosity. Right: The $^{12}$CO moment 0 map. The $^{12}$CO was integrated along channels from -5 to 17 km/s with respect to the rest frame of the source.}
    \label{fig: class0}
\end{figure*}

In addition to the four binary sources that were identified, one candidate Class 0 source was also found in the sample. Located in the same pointing as J18275472-0342384, this source is easily detected in the continuum and shows significant $^{12}$CO emission. However, a sub-mm source with these coordinates is not present anywhere in the literature. In optical wavelengths, the Class 0 source seems to be obscured by nebulosity in images of J18275472-0342384. However, Figure \ref{fig: class0} reveals a filamentary bridge-like structure of $^{12}$CO connecting the the two objects, suggesting that these objects formed together and that the Class 0 object is just now emerging from its obscuring envelope.

\section{Disk mass distributions}
\label{sec:distributions}

\subsection{Intra-region Comparison}

\begin{figure}[htb!]
    \centering
    \includegraphics[width=\columnwidth]{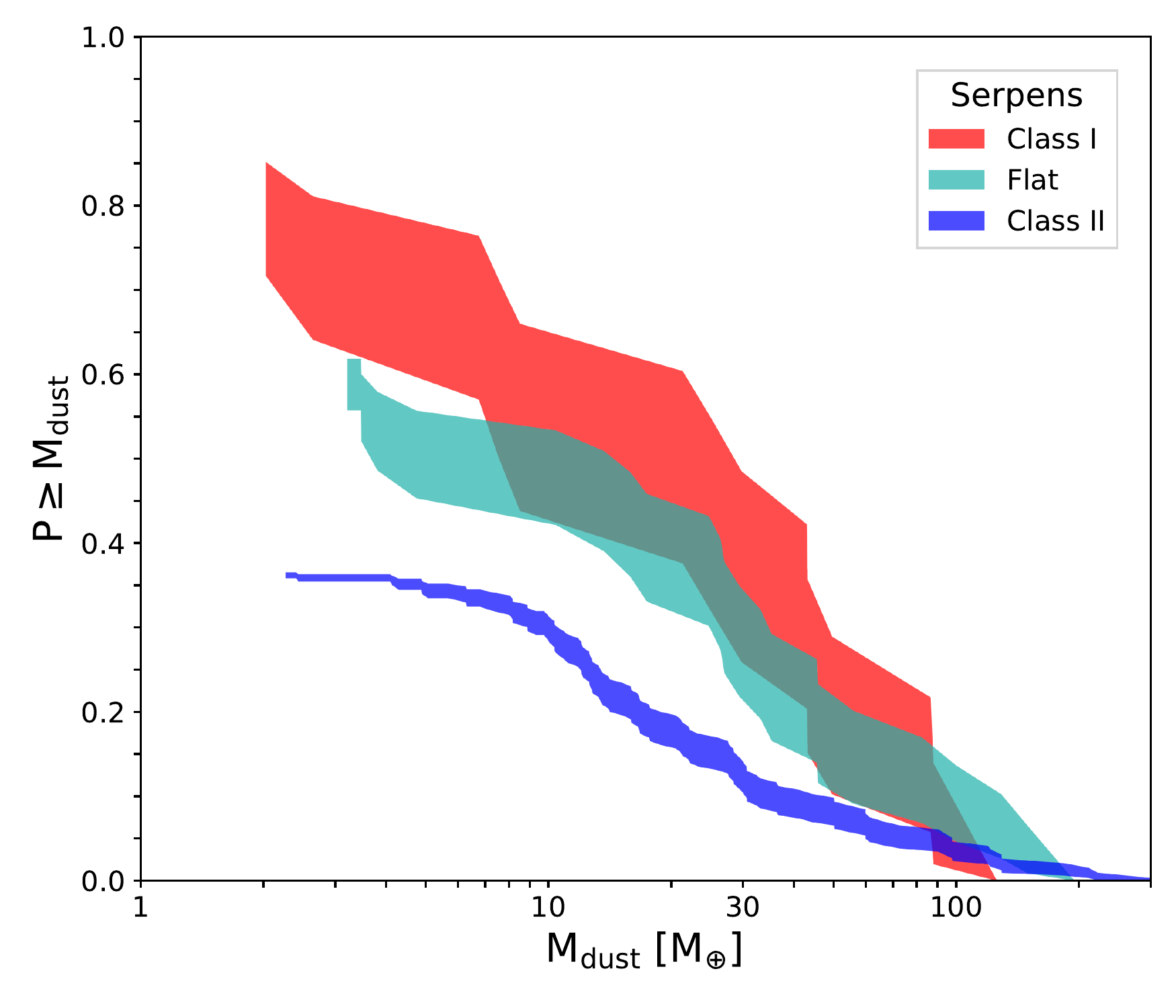}
    \caption{The cumulative dust mass distribution for Serpens disks around YSOs in different evolutionary stages. The shading represents the 68\% confidence interval. For the Class II distribution, the small number of low mass detections below the typical detection threshold produces the flattened region below $10\,M_\Earth$.}
    \label{fig: serpens all class}
\end{figure}

As disks age and YSOs evolve, disk dust masses are expected to decrease due to a combination of accretion onto the star, photoevaporation, and planet formation. There are 16 Class I, 35 Flat spectrum, and 235 Class II sources that we can use to study evolutionary effects on disk masses in our dataset. We used the Kaplan-Meier estimator within the python {\tt lifelines} \citep{Lifelines_2019} package to incorporate the information from the upper limits when creating cumulative mass distributions for each evolutionary group. The distributions for Class I, Flat, and II objects are shown in Figure \ref{fig: serpens all class} and demonstrate a clear progression from high to low mass as the inner disk evolves (as characterized by the infrared slope). The uncertainties are large for Class I and Flat object distributions due to the relatively small sample sizes but the detection rates are reasonably high (87\% and 63\%, respectively). The Class II distribution is better characterized due to the large sample but the detection rate is only 37\%. The flattening of the Class II curve below $\sim 10\,M_\oplus$ is due to a small number of detected sources with masses lower than most non-detection's mass upper limits. This is simply a signature of the relatively wide range in mass sensitivity due to the range of noise in the observations and distances to the targets. However, the low detection rate prevents us from a detailed characterization of the sample by, e.g., fitting a gaussian probability distribution function as in \citet{Williams_2019}.

Above $10\,M_\oplus$, well beyond the median upper mass limit, the distributions are reasonably complete and can be compared. For this subset, the mean masses are 57 $M_\Earth$ for Class I, 57 $M_\Earth$ for Flat, and 48 $M_\Earth$ for Class II. 

The relationship between disk dust mass and stellar mass was also investigated, as it has been previously noted that the two positively correlate \citep{Andrews_2013}. However, only a small fraction of our sources ($\sim$30) had reliable stellar mass measurements. We tried fitting the spectral energy distributions of our sources to various grid models to estimate stellar mass, but the number of degeneracies in our input parameters (spectral type, stellar luminosity, etc.) prevented us from doing so. Nevertheless, we analyzed the $M_{dust}-M_{\star}$ relation for the few sources in our sample with known $M_{\star}$ values and plotted our results in Figure \ref{fig:disk_vs_stellar}. Our results showed a slight positive correlation between $M_{dust}$ and $M_{\star}$, but with a large dispersion, similar to \citet[][Figure 5]{Law_2017}. 

\subsection{Class II Regional Comparison}
\label{sec: class ii}

Class II YSOs have much longer lifetimes, on average, than Class I and Flat spectrum protostars \citep{Dunham_2015} so they provide both greater numbers and a longer temporal baseline to study disk evolution.
Many millimeter wavelength surveys of Class II ``protoplanetary'' disks have been carried out both pre- and post-ALMA and generally demonstrate a monotonic decrease in dust mass from $\sim 1$ to $\sim 10$\,Myr \citep[e.g.,][]{Ansdell_2017, Villenave_2021}.
A notable exception is Ophiuchus, which stands out for its systematically low disk masses despite its young age \citep{Williams_2019}. A possibly similar region is Corona Australis \citep{Cazzoletti_2019}, though new \textit{Gaia} measurements have expanded the membership considerably and suggest a more evolved state \citep{Galli_2020}.
However, Serpens has a similar proportion of Class I, Flat and II sources as Ophiuchus and therefore provides a useful test of how unusual Ophiuchus might be.

\begin{figure}[htb!]
    \centering
    \includegraphics[width=\columnwidth]{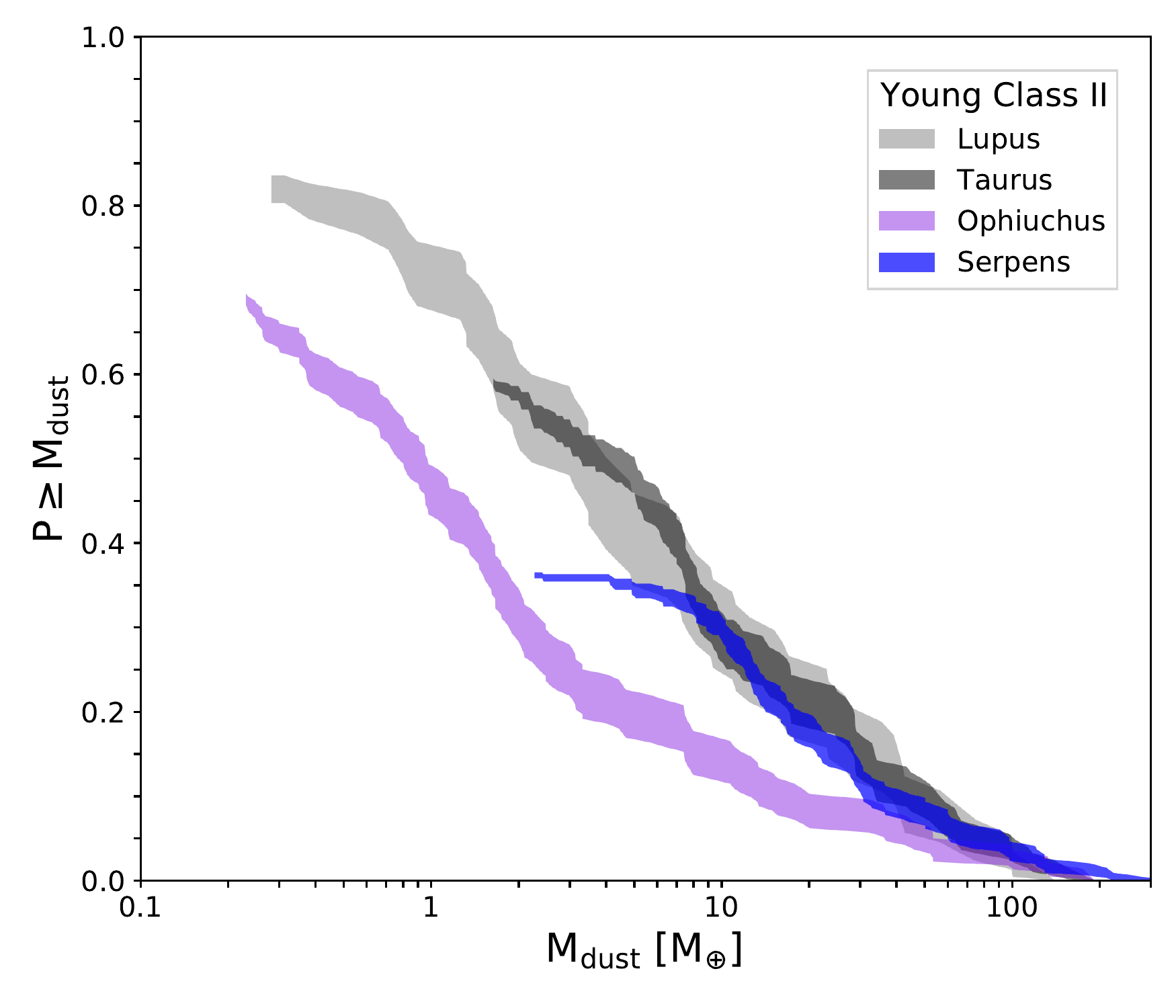}
    \caption{Cumulative dust mass distributions for disks around Class II YSOs in young star forming regions. The Serpens sample is substantially less sensitive than the others and a small number of low mass detections below the typical detection threshold produces the flattened region below $10\,M_\Earth$.}
    \label{fig: class ii}
\end{figure}

Rather than compare the Serpens Class II mass distribution with the many others measured in other star-forming regions, we restrict our attention here to young regions and Ophiuchus in particular. These regions have an average optical YSO age estimated from pre-main sequence tracks to be $\lesssim 3$\,Myr and evolutionary states characterized from Spitzer surveys.
Figure \ref{fig: class ii} plots the Kaplan Meier estimator for Class II sources with masses derived from the same Beckwith $\kappa_\nu$ prescription and uniform $T_{\rm dust}=20$\,K in Taurus \citep{Andrews_2013}, Lupus \citep{Ansdell_2016}, and Ophiuchus \citep{Williams_2019}. 

Although the sensitivity of the Serpens sample is much lower than the other regions, there is a remarkably tight overlap with Lupus and Taurus where the distribution is near complete at masses $\gtrsim 10\,M_\oplus$. This agrees with the results of \citet{Law_2017}, which saw the same behavior within the same mass range when observing with the SMA, but for a much smaller sample. Given the relatively high proportion of embedded protostars, the Serpens region is likely in a slightly younger evolutionary state than Lupus and Taurus as a whole and more similar to Ophiuchus. Nevertheless, Ophiuchus' Class II mass distribution is the anomaly within these comparisons.

\subsection{Class I Regional Comparison}

Similarly to Section \ref{sec: class ii}, a comparison of the disks around Class I and Flat spectrum protostars in different star forming regions provides insight into the earlier stages of disk evolution. As these evolutionary stages are short (0.44 and 0.35 Myr \citep{Evans_2009}, respectively), their numbers are much smaller than those of Class II objects, and Lupus and Taurus do not have sufficient statistics (nor uniform, unbiased surveys) to perform such a comparison. Consequently, we compare Serpens and Ophiuchus with two other regions that are both young and large enough to have many Class I and Flat spectrum sources: Orion and Perseus, surveyed with ALMA by \citet{Tobin_2020} and \citet{Tychoniec_2020}, respectively.

\begin{figure*}[htb!]
    \centering
    \includegraphics[width=\textwidth]{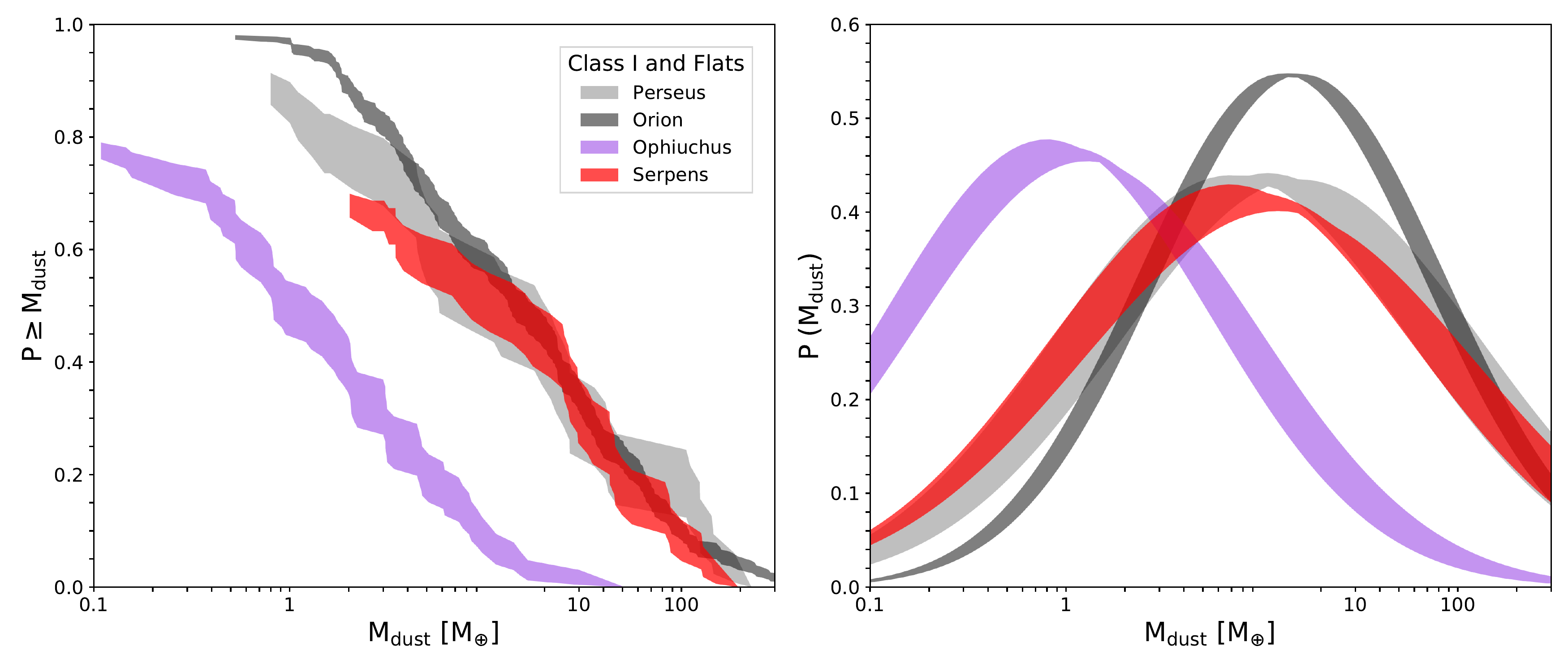}
    \caption{Left: Cumulative dust mass distributions for disks around Class I and Flat spectrum protostars in young star forming regions. Right: Corresponding gaussian distributions for the probability distribution functions of each region. Shading indicates the range of allowed fits.}
    \label{fig: class i}
\end{figure*}

To make the comparison most homogeneous, we recalculated disk masses from the published flux density values using the same $\kappa_\nu$ prescription described in  S\ref{sec:results} and with uniform dust temperatures $T_{\rm dust}=30$\,K for Flat spectrum and $T_{\rm dust}=40$\,K for Class I sources.
To improve the statistics, especially for Serpens, we then bundled the two categories together in the cumulative mass distribution plots in Figure \ref{fig: class i}.

The Serpens distribution is very similar to that of Orion and Perseus but, as for the Class II sources, Ophiuchus is atypical. As the detection rate is higher for these brighter sources in our sample, the cumulative distribution is better characterized, extending to cumulative fractions greater than 50\%. The median disk dust mass values (including upper limits) are roughly 16, 18, 27, and 2 $M_\oplus$ for Orion, Perseus, Serpens, and Ophiuchus, respectively. We note that these values differ from previously published results due to our recalculation of the masses.

With a detection rate of 70\% for the Class I and Flat Spectrum sources, the Serpens distribution is sufficiently complete for the cumulative mass distribution to be fit with an error function and estimate the parameters of a gaussian probability distribution. Indeed, all four distributions in Figure~\ref{fig: class i} are well fit by this method and the results are shown in Table \ref{tab: class i fits}. The means of the distributions are indeed similar for Orion, Perseus and Serpens, whereas Ophiuchus is a factor of 10 lower. Despite this large difference, however, the dispersions are remarkably similar: $\sim 0.7-1$ in $\log_{10} M_{\rm dust}$.

\begin{table}[htb!]
\caption{Gaussian Fits to Class I and Flat Spectrum Disks}
    \centering
    \begin{tabular}{ccc}
    \hline \hline
    Region & Mean $(M_\Earth)$ & $\sigma(\text{log}_{10}(M/M_\Earth))$\\
    \hline
    Perseus & $10.69 \pm 4.40$ & $0.90 \pm 0.01$ \\
    Orion & $14.17 \pm 1.86$ & $0.73 \pm 0.01$ \\
    Ophiuchus & $1.02 \pm 0.24$ & $0.85 \pm 0.02$ \\
    Serpens & $9.31 \pm 2.64$ & $0.95 \pm 0.03$ \\
    \hline
    \end{tabular}
\label{tab: class i fits}
\end{table}

In addition to fitting the observed cumulative dust mass distributions to probability distributions, we tested if the differences in distributions was statistically significant. With the {\tt lifelines} package, we used the log rank test to compare these populations in a non-parametric way that accounted for non-detections.
$p$ values, or the probability that the distributions were different simply because of random chance, were larger than 0.05 when comparing Orion, Perseus, and Serpens with each other, so we cannot claim that these three regions' disks come from statistically different distributions. However, when any of these three regions was compared with Ophiuchus, $p$ values were consistently $<$0.005. Therefore, we can state that the youngest disks in Ophiuchus come from a different parent distribution than those in Orion, Perseus, or Serpens.

\section{Discussion}
\label{sec:discussion}
With one notable exception, the disk mass distributions of the various young star-forming regions analyzed here share striking similarities and largely overlap one another. We have found that the Class II disks in Serpens have essentially the same distribution as in Taurus and Lupus, at least above $10\,M_\odot$ where our survey is near complete.
Moreover, the disks around the embedded Class I and Flat spectrum sources have a significantly similar distribution to those in Orion and Perseus. This is despite the differences in environments: the YSO density in Taurus and Lupus is low ($\sim$1-10 pc$^{-3}$, \citealt{Luhman_2018}) whereas the Serpens sources are strongly clustered,
$\sim$430 pc$^{-2}$ \citealp{Gutermuth_2008b}).

It is important to note that our observations were made at a resolution of roughly 1$\arcsec$, which could blend possible binaries and lead to a disk dust mass distribution that is biased towards massive objects. However, Serpens is a low-mass star forming region, and as stellar multiplicity correlates with stellar mass \citep{Duchene_2013}, we expect this is a minor effect. \citet{Enoch_2011} estimated a protostellar multiplicity fraction of $\sim 10\%$ within the cloud which is too small to significantly affect our conclusions. During the review phase of this manuscript, some higher resolution, $\sim 0\farcs3$ data were taken and are currently being analyzed (Tong et al. in prep). These should resolve tighter binaries concealed within our sample.

Orion is a massive star-forming region with several O stars that produce a strong UV radiation field in their immediate vicinity. This is known to enhance photoevaporation and produce lower disk masses at close range \citep{Mann_2014, Ansdell_2017}, but most disks are many parcsecs away from these stars and, on the scale of the entire star forming region, there is no discernable effect on the disk population when compared to Serpens.
Even though the YSOs in Serpens South are spaced by a typical protostellar core diameter (0.1\,pc, \citealp{Gutermuth_2008b}), the much smaller disks that are rapidly evolving in their centers are not noticeably affected by the crowding.

The notable exception is Ophiuchus, which stands out as anomalous in both the protoplanetary and protostellar disk mass distribution comparisons. It is clearly a young star-forming region as the surrounding molecular cloud has not substantially dispersed, there is a relatively high fraction of Class I and Flat sources \citep{Evans_2009, Dunham_2014}, and the Class II YSOs have high luminosities and ages estimated from pre-main sequence tracks of $\sim 0.1-1$\,Myr \citep{Luhman_1999}. Although there are fewer Class 0 objects in Ophiuchus compared to the other young regions here, such sources are very short-lived and intrinsically rare. The similar proportion of (longer-lived and therefore more common) Class I and Flat objects when compared with Class II objects \citep{Dunham_2015} is a more robust indication that these regions are in similar evolutionary stages and can be equitably compared.

The low disk masses in Ophiuchus were already noted by \citet{Williams_2019} for Class II sources and by \citet{Tobin_2020} and \citet{Tychoniec_2020} for Class I and Flat spectrum sources, respectively. The comparison here with Serpens, which is also a low mass, clustered star-forming region with similar multiple indicators of youth, allows a direct comparison across both protoplanetary and embedded disks and confirms that Ophiuchus is an outlier.
The median mass difference is very large for each disk subset -- about an order of magnitude. Despite \citet{McClure_2010} showing a lower contribution from the envelopes around many Class I and Flat objects, the stark contrast between median masses in Ophiuchus and other regions cannot be explained away by possible source mis-classification.
The low masses at all evolutionary stages suggest that the differences are intrinsic and likely began at formation
\citep[e.g., see][for evidence for a global event]{Forbes_2021}.
Potential avenues to better understand the cause are observational studies of Class 0 sources and cloud-to-disk scale numerical simulations of star formation \citep{Bate_2018}.

\section{Summary}
\label{sec:summary}

We surveyed 320 protoplanetary disks (16 Class I, 35 Flat spectrum, 235 Class II, and 16 Class III) in the Serpens star-forming region with ALMA at 1.3\,mm (Band 6) and analyzed the continuum and $^{12}$CO line data. Our main results are as follows:

\begin{itemize}
    \item We detected 130 sources in the continuum. The derived disk dust masses systematically decline from Class I to Flat spectrum to Class II sources.
    \item We compared the Class II mass distribution with those of other young star-forming regions with large numbers of YSOs that have been surveyed at high sensitivity. Serpens is very similar to Lupus and Taurus for masses greater than the median detection limit $\sim 10\,M_\Earth$. This indicates that the YSO stellar density does not greatly affect the disk demographics. However, the Ophiuchus distribution is shifted to much lower masses.
    \item We compared the mass distribution of Class I and Flat spectrum sources with other young regions where there are near-complete ALMA surveys of their embedded disk population. The Serpens distribution overlaps with those of Perseus and Orion indicating that massive stars do not have a strong effect on the overall disk population. Similar to other studies, we find that Ophiuchus is the outlier with anomalously low masses, and confirm this finding with the log rank test.
    \item Serpens has many similar properties to Ophiuchus as each are very young, low mass, clustered star-forming regions with similar proportions of Class I, Flat spectrum and Class II sources. The direct comparisons shown here reinforce previous conclusions that Ophiuchus disks have anomalously low masses at all evolutionary stages. The mean difference is about an order of magnitude for both Class II and Class I / Flat which suggests that this is a characteristic property of the region that may have been inherited from its initial conditions.
    \item The survey was too shallow and the resolution too low to unambiguously detect CO line emission from the disks, but we were able to identify several protostellar outflows and catalog them in the Appendix. This provides a relatively unbiased sample that would be interesting to follow up with deeper observations and analysis.
\end{itemize}


\acknowledgements
We thank Jason Hinkle and Ruihan Zhang for their comments on this work.
J.P.W. acknowledges support from NSF grant AST-1907486.
N.M. acknowledges support from the Banting Postdoctoral Fellowships program, administered by the Government of Canada. 
C.J.L. acknowledges funding from the National Science Foundation Graduate Research Fellowship under Grant No. DGE1745303.

This paper makes use of the following ALMA data: ADS/JAO.ALMA\#2019.1.00218.S. ALMA is a partnership of ESO (representing its member states), NSF (USA) and NINS (Japan), together with NRC (Canada), MOST and ASIAA (Taiwan), and KASI (Republic of Korea), in cooperation with the Republic of Chile. The Joint ALMA Observatory is operated by ESO, AUI/NRAO and NAOJ. The National Radio Astronomy Observatory is a facility of the National Science Foundation operated under cooperative agreement by Associated Universities, Inc.

This work has made use of data from the European Space Agency (ESA) mission
{\it Gaia} (\url{https://www.cosmos.esa.int/gaia}), processed by the {\it Gaia}
Data Processing and Analysis Consortium (DPAC,
\url{https://www.cosmos.esa.int/web/gaia/dpac/consortium}). Funding for the DPAC
has been provided by national institutions, in particular the institutions
participating in the {\it Gaia} Multilateral Agreement.

\facility{ALMA}

\software{{\tt AstroPy} \citep{Astropy_2013}, {\tt CASA} \citep{McMullin_2007}, {\tt lifelines} \citep{Lifelines_2019}, {\tt matplotlib} \citep{Hunter_2007}, {\tt NumPy} \citep{Harris_2020}, {\tt pandas} \citep{Pandas_2010}, {\tt SciPy} \citep{SciPy_2020}}

\bibliography{bibliography22}{}
\bibliographystyle{aasjournal}

\appendix
\section{Protostellar outflows}
\label{sec:outflows}

\begin{figure*}[htb!]
    \centering
    \includegraphics[width=\textwidth]{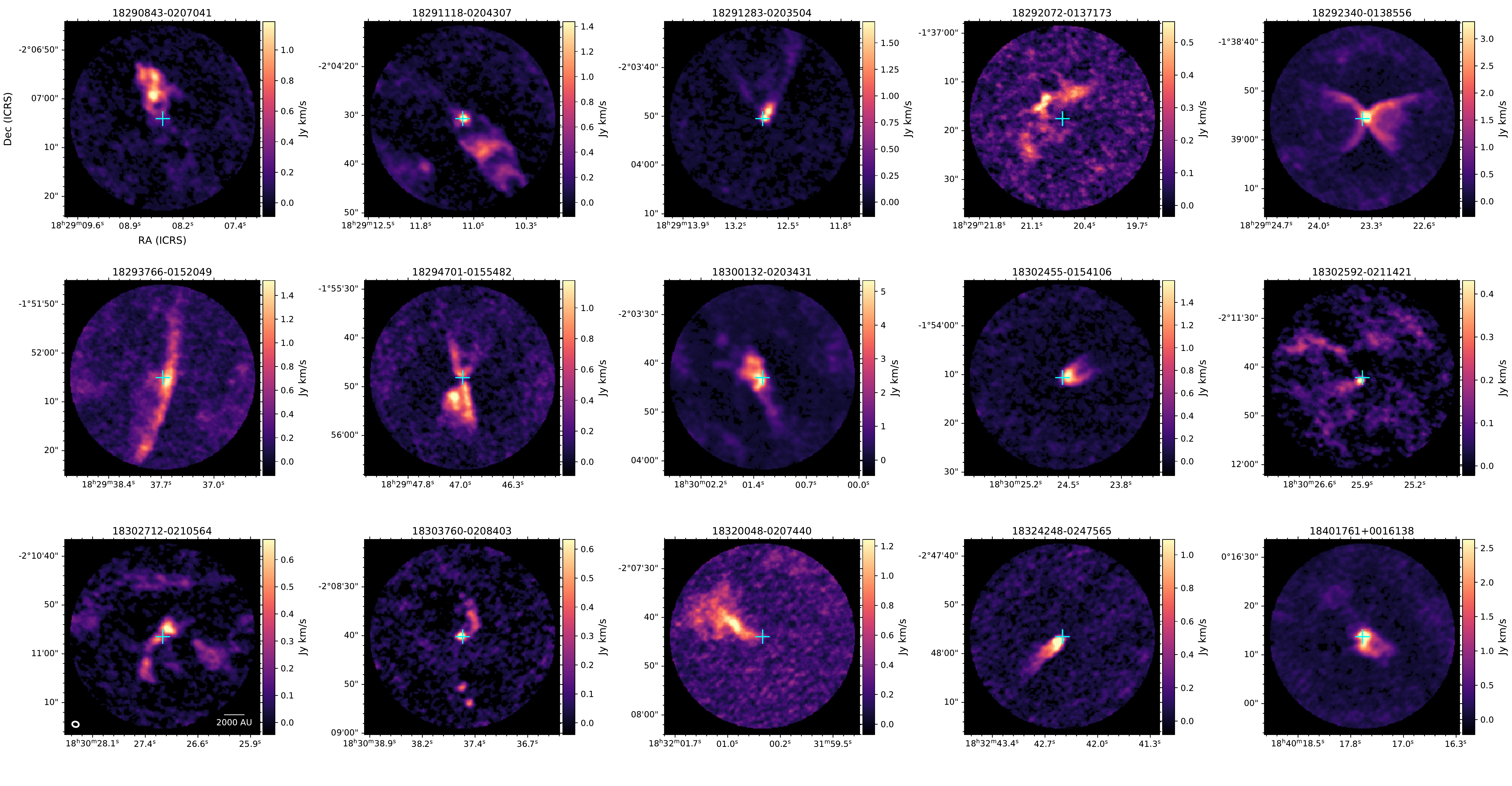}
    \caption{$^{12}$CO integrated intensity maps for eight of our sources. Molecular outflows can be seen emanating from the central protostellar source. A representative beam size and a scale bar showing 2000 AU can be seen in the lower left panel. The location of the sub-mm source is marked with a cyan cross.}
    \label{fig: outflows}
\end{figure*}

Young stars are the drivers of protostellar outflows, or bipolar jets ejecting mass from a YSO \citep{Arce_2007}. These outflows sweep up material nearby, and may therefore influence the formation and evolution of planet-forming disks. Additionally, as outflows are more energetic than the ambient medium and therefore move at distinct velocities, they can be identified separately from surrounding cloud material by manually inspecting channel maps. 
Fifteen molecular outflows originating from the sources in our sample were identified in $^{12}$CO, and shown in Figure \ref{fig: outflows}. Four of the outflows were associated with Class I sources, seven with Flats, and four with Class IIs. Therefore, roughly 20\% of Class I or Flat sources seem to be producing protostellar outflows, while only $\sim$ 2\% of Class II sources do, showing that outflows in Serpens seem to dissipate over time. 

Generally speaking, within our sample, outflows tended to emanate from young sources with strong millimeter detections and inferred dust masses of hundreds of $M_\Earth$.
However, sources J18291118-0204307, J18292072-0137173, J18293766-0152049, J18302712-0210564, and J18320048-0207440
are not detected in the continuum and have disk dust mass upper limits on the order of only a few $M_\Earth$. They clearly are visible in $^{12}$CO and have the characteristic collimation of an outflow, as well as emission in the specific velocity range of other outflows from detected sources in our survey. Deeper observations will be required to learn more about their properties.

\section{Disk and Stellar Mass}

\begin{figure}
    \centering
    \includegraphics[width = \columnwidth]{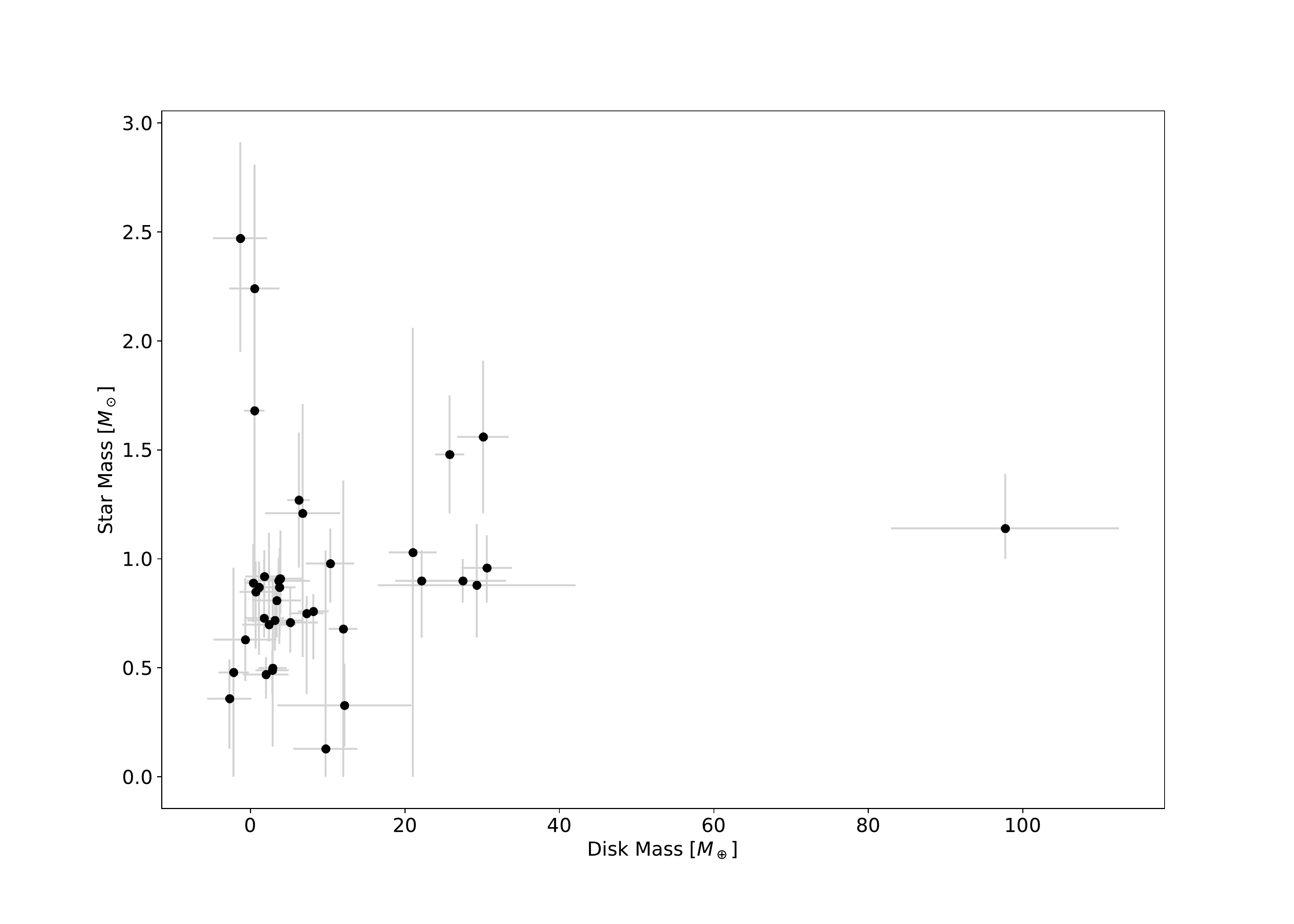}
    \caption{Stellar mass as a function of disk dust mass. A correlation is seen between the two quantities, but with a high degree of scatter, similarly to what was found in \citet[][Figure 5]{Law_2017}.}
    \label{fig:disk_vs_stellar}
\end{figure}

\end{document}